%% file: pp.tex
\documentstyle{europhys}
\input euromacr

\newcommand{\rem}[1]{}
\newcommand{\Kpp}{{\Gamma^{\parallel,\perp}}}
\newcommand {\Amat}{{\bf A}(\qq)}
\newcommand {\Bmat}{{\bf B}(\qq)}
\newcommand {\Cmat}{{\bf C}(\qq)}
\newcommand {\Gpar}{{{\bf G}^\parallel}}
\newcommand {\Gperp}{{{\bf G}^\perp}}
\newcommand {\qq}{{\bf q}}

\begin{document}

\euro{}{}{}{}
\Date{}

\shorttitle{W.-J. ZHU et al: PHONON-PHASON COUPLING ETC.}
%
\title{Phonon-Phason coupling in icosahedral quasicrystals}
\author{W.-J.~Zhu\footnote{Present address: IBM T.J. Watson Research Center  
              P.O. Box 218, Yorktown Heights, NY 10598}
              and C.~L.~Henley}
\institute{Dept. of Physics, Cornell University, Ithaca NY 14853-2501}

\rec{}{}

%
\pacs{
\Pacs{61}{44Br}{Quasicrystals}
\Pacs{62}{20Dc}{Elasticity, elastic constants}
\Pacs{61}{10Dp}{Theories of X-ray diffraction and scattering}
\Pacs{71}{15Pd}{ Molecular dynamics calculations (Car-Parrinello) 
and other numerical simulations}
      }

\maketitle

\begin{abstract}
From relaxation simulations of decoration-based 
quasicrystal structure models using microscopically based
interatomic pair potentials, we have calculated the 
(usually neglected) phonon-phason coupling constant. 
Its sign is opposite for the two alloys studied, 
$i$-AlMn and $i$-(Al,Cu)Li; a dimensionless measure of 
its magnitude relative to the phonon and phason elastic 
constants is of order $1/10$,  suggesting its effects are
small but detectable. 
We also give a criterion for when phonon-phason effects are
noticeable in diffuse tails of Bragg peaks. 
\end{abstract}



The elastic description of an icosahedral quasicrystal 
includes both ``phonon'' (ordinary) strain and ``phason'' strain. 
``Phason'' strain parametrizes the way in which the local tile 
distribution deviates from icosahedral symmetry, as forced by
the constraint of packing with surrounding tiles. 
Correspondingly there are two easily measured ``phonon'' elastic
constants (ordinary bulk and shear modulus), and two ``phason'' elastic
constants. The latter have been computed theoretically for various 
tilings~\cite{tang,shawet91,newhen95,mihen98}
and recently measured experimentally \cite{boi95}. 

There remains a cross-term with one elastic constant, 
the ``phonon-phason'' coupling $\Kpp$, which has not previously been
measured experimentally or theoretically.  The constant $\Kpp$ enters
the shape of the diffuse tails around Bragg peaks \cite{jarne88}, can
drive a quasicrystal phase unstable with respect to a related crystal
phase \cite{wid91,ish92}, and affects the strain fields of dislocations in
quasicrystals, which have been studied by diffraction contrast in
transmission electron microscopy \cite{woll97,yang98}. 
Previous authors \cite{boi95,jarne88,yang98,bactre98}
simply assumed $\Kpp$ was small compared with the other elastic constants. 

The present work is the first attempt to compute $\Kpp$ theoretically.
As explained below, 
we extract it numerically from simulations
in which we relax the atomic positions of a finite 
``periodic approximant'' of the quasicrystal and 
measure its spontaneous shear distortion in response to the
its intrinsic {\it phason} strain, for several different approximants. 
This was carried out 
for moderately realistic models of both $i$-AlMn and
$i$-(Al,Cu)Li (modeled as a pseudo-binary), 
representing the two major classes of quasicrystal. 
(Here ``model'' denotes a combination of a tiling, 
an atomic decoration rule, and a microscopically based pair potential.)
In our conclusion, we crudely estimate ratios which determine
the experimental observability of $\Kpp$. 


\section{Elastic theory}
This is formulated in terms 
of two kinds of strain fields:
the ``phonon'' (ordinary) strain tensor has
components $u_{ij}=(\partial_i u_j + \partial_j u_i)/2 $ 
where {\bf u(r)} is the phonon displacement.  
The phason strain components are $v_{ij}=\partial_i v_j$. 
Here {\bf v(r)} is the phason (also called ``perp'')
displacement field, defined for quasicrystals and/or tilings; it
can be constructed for any configuration of the tiles~\cite{hen91ART,el85},
and parameterizes the local imbalance in tiles of
different orientations. Periodic approximant structures
\cite{elhen85} of the quasicrystalline state necessarily acquire a
non-zero phason strain.  

Group theory \cite{jarne88,lub88} permits the
following terms in the elastic 
free energy (in the notation of \cite{wid91}, and adopting the
summation convention):
\begin{eqnarray}
\label{eq-elastic}
  F &= &F_{phonon} + F_{phason} + F_{phonon-phason} \\
  F_{phonon} &= & \int d^3{\bf r} 
       \frac{1}{2} \left[\lambda u_{ii}^2 + \mu u_{ij}u_{ij}\right] 
       \nonumber \\
  F_{phason}& = & 
        \int d^3{\bf r} 
          \frac{1}{2} \left[ K_1 v_{ij}v_{ij} +
	K_2 \left\{ v_{kk}^2 - \frac{4}{3}v_{ij}u_{ij} +\left[(\tau
	v_{12} + \tau^{-1} v_{21})^2 + {\rm perms} \right]\right\} \right] 
        \nonumber \\
  F_{phonon-phason} &= &\int d^3{\bf r} 
      \left. \Kpp \left\{ \left[ v_{11}(u_{11} - \tau u_{22} + \tau^{-1} u_{33})
     + 2 u_{23}(\tau^{-1} v_{23} - \tau v_{32})\right] + 
     {\rm perms} \right\} \right.
         \nonumber
\end{eqnarray}
Here ``perms'' represents two terms 
obtained from the preceding term (within the same parenthesis) 
by cyclic permutations of the indices (123). 
The term $F_{phonon}$ is the free energy of any isotropic
harmonic solid with strain. It has two independent Lam\'e constants 
$\lambda$ and $\mu$, the latter being the shear modulus.
Its ``phason'' analog is the $F_{phason}$ term
with elastic constants $K_1$ and $K_2$.
The $F_{phonon-phason}$ term  couples the shear strain with 
one representation of the phason strain.

As in ordinary crystals, the elastic free energy
describes the effects of thermal fluctuations on diffraction peaks, 
as well as possible soft-mode instabilities of the quasicrystalline state.
\rem{Among this is a phasonic Debye-Waller factor.
Possible modes of instabilities due to a phason-phonon coupling
 include modulated phase, an approximant or a crystal structure
\cite{hen91ART}.}

One application of elastic theory is to dislocations in
a quasicrystal, since any Burgers vector necessarily has both phonon and
phason components.  
The smaller the ratio $K_1/\mu$, 
the smaller the (expected) phonon component and the larger
the  phason component for a stable dislocation~\cite{woll97,yang98}. 
This might be modified by a nonzero $\Kpp$.

\rem {CLH's recollection was that
the role of $\Kpp$ was emphasized in the earliest papers, 
namely Widom 1989.   But I didn't find that on rereading.
My own paper (Mexico conference) is noteworthy mainly for
an argument -- probably invalid -- that the phonon-phason 
coupling was small.}

\section{Method for measuring phonon-phason coupling $\Kpp$}

Given any structure model, 
$\Kpp$ may be measured from the intrinsic distortion
common to all approximant tilings.
Its origin is the (known)  nonzero uniform {\it phason} 
strain $(v_{ij})$
characteristic of each periodic crystalline approximant.
\rem {The presence of the $F_{phonon-phason}$ term allows a phason response
(tile re-arrangement) to a phonon strain, as well as a phonon response
(shear or distortion) to a phason strain.
At sufficiently low temperatures, the phason
fluctuation freezes out \cite{hen91ART} and tile flips become
infrequent; thus the $F_{phason}$ part of the free energy may be
ignored, leaving a Hamiltonian with two remaining parts.  The
$F_{phonon-phason}$ term remains non-zero when both phonon and phason
strains are anisotropic.}  
Given $(v_{ij})$, the elastic theory then predicts a corresponding nonzero
uniform (phonon) shear strain. Concretely, this means
each tile spontaneously distorts a bit proportional to 
$\Kpp$ times $(v_{ij})$, so that tile edges in different
directions are no longer quite related by icosahedral symmetry.  

Consider in particular an orthorhombic (pseudo-tetragonal)
approximant with $v_{11}=v_{22} \neq v_{33}$, and all other
$v_{ij} =0$. 
We allow it to distort with fixed-volume shears
$u_{11}=u_{22}=-\epsilon/2, u_{33}=\epsilon$, 
so that the bulk modulus $\lambda$ does not contribute to the energy. 
Under this constraint, the minimum free energy occurs at
   \begin{equation}
   \label{eq-kkk}
   \epsilon = \epsilon_{min} \equiv  -  \frac{\Kpp}{2 \mu} {  } 
(v_{33}-v_{11})
   \end{equation}
Given a particular approximant, we can numerically compute 
the relaxed structural energy difference $E_{rel}(\epsilon)$
for several choices of shear magnitude $\epsilon$.
Obtaining $\epsilon_{min}$ and $\mu$ by fitting 
   \begin{equation}
   E_{\rm rel}(\epsilon) =
   {1 \over 2} \mu (\epsilon-\epsilon_{min})^2 +{\rm const},
   \label{eq-Erel}
   \end{equation}
we finally deduce $\Kpp$ using (\ref{eq-kkk}).

\rem{We had tried alternatively to use Fourier Space
formulations, or to seek relationship between the relaxed positions
and their phason coordinates.  None of these attempts provide a
sensible explanation that would show phason strain dependence.}

\rem {WJZ's remark. 
To further visualize this scenario, we can consider the extreme case
of having a periodic tiling of only prolate rhombohedra, pointing the
same way, to represent a large phason strain with imbalance tiles.
If the quasicrystalline phase is highly preferred, there must be a
large amount of frustration present in the atomic structure of the
periodic tiling.  The heavily strained tiles would distort
proportional to $\Kpp$.}

\section {Specification of the models}
Even a crude estimate of the phonon-phason coupling 
requires sensible atomic models  and
realistic interatomic potentials \cite{marek1,marek2},
otherwise the structure is not very stable and likely to
rearrange irreversibly after small perturbations.
There are two important and different icosahedral structure 
types: the Frank-Kasper class and the Al-transition metal class, 
represented by $i$-(Al,Cu)Li and $i$-AlMn. 
We begin with the specification of the structural model, 
the first kind of input to our calculation. 

The atomic positions in the unit cell to be relaxed 
are determined by three partly independent structural factors: 
tiling, decoration, and approximant.~\cite{marek1}.  
The simplest icosahedral tiling geometry is the well-known
three-dimensional rhombohedral tiling (3DRT)
with edges along 5-fold symmetry axes~\cite{elhen85,henel86}.
\rem {However, decorations on this tiling
tend to encounter steric conflicts along the short diagonal of the 
oblate rhombohedra.
The problematic occurrences of oblate rhombohedra are minimized
in the CCT, which also maximizes the  presence of clusters,} 
The Canonical Cell Tiling (CCT) \cite{hen91CCT} 
is an alternative tiling of 4 larger cells, which 
permits adjustments of the decoration to eliminate 
high-energy environments. 

In the AlMn system we have used
one a representative decoration rule 
from each of three families of decoration rules for the 
CCT ~\cite{marek1,marek2}
These are labelled DF, DS, and LS in the scheme of \cite{marek1}, 
according to 
their tendency to dense (D) or loose (L) space-filling by the atoms, 
and their space-group symmetry (S=Simple Icosahedral or
F=Face-Centered Icosahedral). 
~\footnote{
We used variants DF1.1, DS1.1, and LS1.1; we have not
tried the other variants possible within each family, 
which differ by changes of the chemistry or density on 
particular sites~\cite{marek1,marek2};
a representative LF1.1 from the fourth family of decorations
in \cite{marek1} 
did not even show a quadratic $E_{\rm rel}(\epsilon)$ function.}
In the (Al,Cu)Li system.
we only use the ``Henley-Elser'' (HE) decoration
of the 3DRT~\cite{henel86} 
the version in which rhombohedra are
grouped into {\it disjoint} rhombic dodecahedra wherever possible). 
The HE decoration may also be applied to 
the CCT (which can be decomposed into a  special case of the 3DRT
\cite{hen91CCT}).

Having chosen the tiling type, we may select among the available
anisotropic approximants.  The lattice parameters for the unit cell of
quasicrystal approximants are constrained to have only lengths which
are powers of $\tau$ times the quasi-lattice parameter \cite{elhen85}.
We simplify the conventional notation ``$p/q$'' \cite{elhen85} for a
rational approximant of $\tau$ by only using ``$p$'' in our labelling.
Thus, the anisotropic approximants are, ordered by increasing system
size, 112, 221, 223, 331, 332, ... \cite{shawet91,oh89}.  In
the CCT case, some of these cannot be tiled by the 4 CCT cells;
in particular, there is no approximant between 331 and 335~\cite{newhen95}.

Phason strain will be measured in units where the
the rhombohedron edge length $a_R$ is taken to be unity in real space, 
as well as the corresponding edge in ``perpendicular'' space~\cite{el85}. 
Thus $v_{\alpha\alpha} = \tau^{-3}, -\tau^{-4}, \tau^{-5}$
when the $\alpha$ entry in the approximant's label is 
1,2, and 3 respectively. 

The second input in the specification of a model is
the choice of pair potential. 
This is the only feasible approach to the energetics, 
since the system size of atomic structures in typical
quasicrystal approximants is orders of magnitude larger than the
maximum allowed for {\em ab initio} calculations.

We adopted available well-tested potentials,
constructed using linear response theory \cite{haf89},
for the binary AlMn \cite{marek2} and the pseudo-binary (Al,Cu)Li
\cite{winet94} systems.~\footnote{
Neither of these alloys is an equilibrium quasicrystal;
the potentials for better quasicrystals were not yet available.}
The AlMn pair potentials involve the
conventional perturbation theory for the nearly-free electrons in the
Al-Al interaction, and Green's function treatment of Mn $d$-bands
\cite{zou93}.  
For the (Al,Cu)Li system, ternary potentials were first constructed
from pseudopotential-based perturbation theory and a tight-binding
approximation of the Cu $d$-orbitals, and then the
pseudo-binary potentials were obtained through proper averaging to
reflect the chemical disorder in the system \cite{winet94}.
We call such pair potentials 
``realistically oscillating'' \cite{marek2} 
because they feature Friedel oscillations out to several Angstroms.  
These potentials were tested in
previous studies (see \cite{marek2} and \cite{winet94});
as a further check we applied the (Al,Cu)Li
potentials to various crystal structures related to the
Frank-Kasper family, and reproduced a sensible phase diagram
\cite{wjz96}.

\section{Calculations and results}
We first constructed the atomic structure for each combination 
of model and pseudo-tetragonal approximant shown in table I. 
(Its first three columns identify the alloy and potential, the tiling
type (and decoration), and the approximant, using notations explained above. 
The next larger approximant would have be 
``335'' which was too large for our relaxations.

For our selected pair potential, we always chose
a cutoff slightly beyond $6 \AA$, so as to sufficiently include
the first two potential wells \cite{marek2}.  
We applied to the cell a uniform fixed-volume shear $\epsilon$ 
and then relaxed the positions (keeping the new cell shape fixed.) 
A modified conjugate gradient energy
relaxation for the system converges to within $10^{-6}$ eV/atom. 
We sampled a grid of $\epsilon$ values to derive the
relaxed energy curve $E_{\rm rel}(\epsilon)$
\rem{We did include the much larger ``335'' approximant,
featuring thousands of atoms, in unrelaxed energy estimations.
The  ``335'' approximant became a practical limit (in size) to
what we can use and meaningfully calculate $\Kpp$: not only is
 the energy relaxation prolonged by the system size,
but more seriously the optimal (minimum energy)
shear $\epsilon_{min}$ is approaching zero quickly, requiring far
more sampling at closer intervals of shear, and produces results that
have a large percentage error.}
which was fitted to (\ref{eq-Erel}).  Combining this with 
(\ref{eq-kkk}) we extracted 
the shear modulus $\mu$ and  phonon-phason coupling $\Kpp$, expressed in
table I in units of eV/atom. 

\begin{table}
\centering
\caption{Computed optimal shear $\epsilon_{min}$, with
shear modulus $\mu$ and
phonon-phason coupling $\Kpp$ in eV/atom. Text explains first three columns.} 
\begin{tabular}{llcccc} 
\hline
Potential &Structure &Approximant &$\epsilon_{\rm min}$ &$\mu$ 	&$\Kpp$	
\\ 
\hline
AlMn   &CCT(DF)	&3{ }3{ }1 &{ }0.03\%   &6.7	&{ }0.021 
\\ &CCT(DS)	&3{ }3{ }1	&{ }0.11\%	&6.3	&{ }0.066	
\\ 
	&CCT(LS) &3{ }3{ }1	&{ }0.22\%	&7.6	&{ }0.168	
\\ 
\hline
(Al,Cu)Li   &CCT(HE) &3{ }3{ }1	&$-1.00$\%	&1.9	&-0.187	
\\ \cline{2-6}
	&3DRT(HE) &1{ }1{ }2	&{ }1.13\%	&2.1	&-0.146	
\\ 
	&	&2{ }2{ }1	&$-0.81$\%	&2.2	&-0.110	
\\ 
	&	&2{ }2{ }3	&$-0.16$\%	&2.4	&-0.063	
\\ 
	&	&3{ }3{ }1	&$-0.80$\%	&2.2	&-0.177	
\\ 
	&	&3{ }3{ }2	&{ }0.16\%	&1.8	&-0.047	
\\ \hline
\end{tabular}
\label{t-result}
\end{table}

For a given interatomic potential, the ordinary shear stiffness
has reasonably consistent values.
 From table I, we first note $\mu \approx 7$ eV/atom 
in $i$-AlMn or $2$ eV/atom for $i$-(Al,Cu)Li. 
\footnote {
Note that $i$-AlMn has a larger $\mu$ because
its Al-Al potential has
an energy scale about twice as big as in the (Al,Cu)Li case; 
in turn, this is due to the different electron densities.}
The number density is about $6.5$  atoms$/{a_R}^3$ in $i$-AlMn\cite{marek1}
or $7.8$ atoms$/{a_R}^3$ in $i$-(Al,Cu)Li\cite{henel86}, with $a_R$ about 
$4.6 \AA$ or $5.05 \AA$, respectively.  This gives the predictions
$\mu \approx  75$ GPa in $i$-AlMn or $20$ GPa in $i$-(Al,Cu)Li, 
to be checked against the experimental values 
$\mu = 65$ GPa in $i$-AlPdMn~\cite{amazit} or $41$ GPa in 
$i$-AlCuLi~\cite{spoor}.

Less surely,  the table shows $|\Kpp| \sim 0.1$ eV/atom. 
Although the {\it magnitude} of the phonon-phason coupling is 
ill-determined in table I, its {\it sign} is 
quite clear and {\it not} universal: it is positive for AlMn and
negative for (Al,Cu)Li.  We have no microscopic explanation for
these signs, and so cannot tell whether they hold 
generally for Al-TM and Frank-Kasper type quasicrystals.

The fitted $\Kpp$ for $i$-(Al,Cu)Li from different approximants
varies by a factor $\sim 4$. 
We attribute this to finite-size effects due to the smallness
of the tractable approximants:
the strain relaxation might be especially sensitive
to certain rare environments,  occurring
only one or two (or no) times in various approximants, 
producing big fluctuations in the inferred $\Kpp$. 
\rem {The number of tiles of a given type and orientation 
may differ substantially from the average in a large system.}

We also relaxed all the 3DRT approximants of the (Al,Cu)Li structure 
using Roth's single-well Lennard-Jones (LJ) potential
with non-additive radii~\cite{roth}, rescaled to roughly
match the first well of the realistic potentials. 
When the simulated binary alloy freezes under these LJ potentials, 
it is known to form a disordered quasicrystal 
of the Frank-Kasper class~\cite{roth}. 
We found that $\Kpp$ is much smaller in this case; 
our LJ results are tabulated in \cite{wjzthesis}. 
\rem {As noted elswhere in the text, 
the smallness of $\Kpp$ is somewhat puzzling: 
the single-well potentials had greater
frustration in \cite{marek2}, so we expected them to show
a larger $\Kpp$.}
\par
\rem {Specifically, our LJ 
length unit is rescaled so that the first minimum of the Al-Al
potentials comes at the same radius as in (Al,Cu)Li; our
LJ energy is rescaled so that the well depth of the Li-Li potential matches
that in the realistically oscillating potential, namely
about $0.002 {\rm Ry}$
[as measured by the difference between the first minimum and the
second maximum]. 
Of course, 
Al/Cu corresponds to the ``small'' type atom and
Li to the ``large'' type atom.}

Our calculation depended on the assumption that
the optimal shear is a function only of the approximant's
phason strain, and insensitive to tile rearrangements 
(which do not change phason strain). 
Unfortunately, the only such comparison in our data set is
between the 3DRT and CCT tilings in the (Al,Cu)Li 331 approximant.
In that case (see Table 1), 
similar elastic constants were indeed fitted.

We also adopted a crude and implausible ``frozen'' tiling assumption, 
that the energetic and entropic contributions to the elastic
free energy are independent and linearly additive. 
\rem {We added the free energy of {\it rigid} tiles in a random tiling, 
to the energy of a {\it frozen} (not rearranged) tiling at $T=0$.}
The value of $\Kpp$ might be substantially renormalized
(and be temperature-dependent)
if we allowed atom relaxations {\it and} tile reshufflings simultaneously.

\section {Discussion}
We may now assess whether the phonon-phason term is strong enough
to be measured experimentally, and 
even perhaps to affect the quasicrystal's stability.
Near a Bragg vector $\Gpar$, the diffuse intensity (following
Ishii's notation\cite{ish92}) 
is proportional to
    \begin{equation}
        (\Gperp, \Bmat^{-1} \Gperp)
        - 2 (\Gpar, \Amat^{-1} \Cmat \Bmat^{-1} \Gperp)
        + (\Gpar, \Amat^{-1}  \Gpar)
    \label{eq-diffuse}
    \end {equation}
We wish to estimate the relative importance of these three terms,
representing respectively contributions 
from ordinary ``phonons'', from ``phason'' fluctuations, and from the
phonon-phason cross term. 
The $3\times 3$ matrices $\Amat$, $\Bmat$, and $\Cmat$ 
are derived from the elastic free energy (\ref{eq-elastic}) 
and  their formulas are given in eqs. (2.3)--(2.5) of \cite{ish92}. 
Each matrix is a sum of terms proportional to the corresponding 
elastic constants and homogeneous of second order in $|\qq|$.
Crudely speaking, the matrices scale as
$||\Amat|| \sim \mu |\qq|^2$, 
$||\Bmat|| \sim  K_1 |\qq|^2$, 
and $||\Cmat|| \sim \Kpp |\qq|^2$. 
(This scaling neglects their strong angular 
dependences on $\qq/|\qq|$; it also 
ignores the distinctions
between the two phonon or the two phason elastic constants, 
since indeed $\lambda \approx 2\mu$ in a typical solid, 
and $|K_2|/K_1 =O(1)$ for all cases quoted below.)

We may conveniently express important criteria using ratios which
all have the same dimensions as phason strain. 
First, the phason fluctuations 
-- the first term in (\ref{eq-diffuse}) -- 
surely dominate that sum, 
as assumed in the analysis of \cite{boi95}, thus
  \begin{equation}
     |\Gperp|/|\Gpar | 
     > |\Kpp|/\mu .
  \label{eq-ineqphason}
  \end{equation}
Second, for the phonon-phason coupling to be measurable, it ought to 
be the dominant correction, 
i.e. the second term in (\ref{eq-diffuse}) is the next largest.
Thus
  \begin{equation}
    |\Gperp|/|\Gpar | 
     > K_1/|\Kpp| .
  \label{eq-ineqpp}
  \end{equation}

Adopting the simplest sort of random-tiling scenario~\cite{hen91ART}, 
all of the phason elastic term is part of the
tiling's configurational entropy. 
Simulations find  $K_1/T = 0.81$ 
$a_R^{-3}$
with $K_2/K_2 = 0.61$ for the 3DRT
\cite{tang,shawet91}. 
For a model approximating the CCT~\cite{mihen98}, 
they find 
$K_1/T$ is (at most) $2.5  a_R^{-3}$
with $K_2/K_1 \approx -0.75 $. 
\footnote {
A less reliable calculation,  but for the true CCT, 
found $K_1 \approx a_R^{-3}$.~\cite{newhen95} }
Furthermore, phason fluctuations fall 
out of equilibrium not so far below the melting temperature $T_m$, 
so we will replace $T \to T_m \approx 10^3 {\rm K}$ in these estimates. 
Thus, using the figures mentioned earlier for $a_R$ and the number density, 
a crude estimate for either alloy 
is $K_1 \approx 0.03$ eV/atom,  in the units of table I. 
(The experimental value of $K_1$ in $i$-AlPdMn 
~\cite{letoublon98} seems to be a factor of 10 smaller.)

The above quoted estimates for $\mu$ (for
$i$-AlMn),  $\Kpp$, and 
$K_1$, when inserted into the right-hand sides of (\ref{eq-ineqphason}) 
and (\ref{eq-ineqpp}), give respectively 0.015 and 0.3.
Thus a Bragg peak may be found
\footnote {
Bragg peaks exist with arbitrarily small $|\Gpar|/|\Gperp|$, 
but with exponentially small intensities~\cite{el85}.}
such that the $\Kpp$ term is the first correction. 
This correction would cause differences in shape of the
(otherwise similar) diffuse scattering contours of
two Bragg peaks with $\Gpar$ in the same direction, 
as observed recently \cite{letoublon98}. 

\rem{Dimensionless ratio}

There is an additional correction (see e.g. \cite{ish92}) to every
term of (\ref{eq-diffuse}), scaling with the relative order 
   \begin{equation}
     r\equiv |\Kpp|^2/K_1 \mu 
   \end{equation}
(provided we assume $K_2=O(K_1)$ and $\lambda =O(\mu)$, as
in the previous discussion). 
Evidently $r$ is the (dimensionless) ratio of the right hand sides of 
(\ref{eq-ineqphason}) and (\ref{eq-ineqpp}). 
Using the elastic constants
suggested above for $i$-AlMn and $i$-(Al,Cu)Li gives
$r \simeq 0.05$ and $r \simeq 0.2$, respectively, 
uncertain by an order of magnitude either way. 
This correction provides a second way
that $\Kpp$ affects the diffuse scattering.
\rem{The larger approximants show a significant
decrease in $r$, as a result of the smaller
optimal shear distortion.}
Thus too, a version of $r$ shows up in the criterion
for nondivergence of the fluctuations (and diffuse scattering), 
Widom's eq.~(7):
$ C_\mp (\Kpp)^2/(\lambda + 2\mu)(K_1 \mp 4 K_2/3) < 1$ 
where $C_- = 4/9$ and $C_+=4$. 
\rem{These formulas are less sensitive to the phonon-phason coupling
than Widom's (5) for the uniform instability, 
recalling that $\lambda + 2\mu \approx 4\mu$. 
By the way, in the CCT~\cite{newhen95,mihen98} or
in real $i$-AlPdMn~\cite{boi95,letoublon98}, 
where $K_2/K_1 \approx -0.75 $ or $-0.5$, 
$K_1+4K_2/3$ is small and the equation with $C_+$ 
might be more important than the one with $C_-$.}

The same ratio $r$ also determines the importance of $\Kpp$ in
the onset of instability to a {\it uniform} phason strain, as
discussed in \cite{wid91} and \cite{ish92}. 
Such phason-elastic instabilities might 
generate the approximant phases -- e.g. rhombohedral $R(AlCuFe)$ --
typically occurring for compositions close 
to the quasicrystal's~\cite{ish92}.
The stability criterion is given in Widom's eq. (5) \cite{wid91}:
$ 3 |\Kpp|^2/(K_1- 4 K_2/3) \mu < 1$. 
If $K_2/K_1 \approx 0.6$, as in the random 3DRT~\cite{tang} or
in $i$-AlCuFe~\cite{letoublon98}, the criterion says $r < 0.07$. 
If our estimate of $r$ is valid for real $i$-AlCuFe, then it is
close to a phonon-phason driven instability. 

The smallness of  the phonon-phason constant
has been  rationalized by the notion that it measures the
``frustration'' of a material, defined as the mismatch of the
ideal interatomic spacings and the well 
radii of the potentials~\cite{hen91ART}.
In the unfrustrated case, different terms of the interactions
are independently satisfied by the atoms, which occupy 
tiling-like special positions.  
This permits good quasicrystal order, since 
the tile-tile energy may be low for more than one way
of packing tiles along a face. 
At the same time, the lack of unbalanced forces
on the atoms means there is less tendency for tile shapes to distort
depending on the tile packing. 
In this way the ``binary tiling'', a two-dimensional atomic
toy model~\cite{wid87}, 
has a quite small $\Kpp$~\cite{strand89}. 
\rem {The binary-tiling potentials are constructed so
that all neighbor distances are at the bottom of the
LJ potential wells.  By the way, 
[~\cite{strand89}] suggests $4C^2/KQ$, a dimensionless ratio
analogous to our $r$, becomes large
$(4(3.2)^2/0.6(77.3))\approx0.9$, near the melting temperature.
}

However, our results appear to contradict the above argument:
among the three AlMn decorations of the CCT 331 approximant, 
the ``LS'' decoration is best adapted to the structure~\cite{marek2}, 
yet it shows the {\it highest} $\Kpp$.
(We suppose this is somehow due to the small voids common in 
the ``LS'' structure, which  allow greater atomic rearrangement
in response to phonon or phason shears.)
Similarly, on the 3DRT tilings with the $i$-(Al,Cu)Li 
structure,  the $i$-(Al,Cu)Li potentials in Table I 
give a much larger phonon-phason coupling than the
poorer LJ model (computed in~\cite{wjzthesis}).

In conclusion, we have numerically estimated  the phonon-phason
coupling $\Kpp$ for representatives of both major classes of icosahedral 
quasicrystal.  We found it had a non-universal sign, and
(with very large uncertainties)
it was large enough in magnitude to
be observable and perhaps even to dominate the mechanism of
the quasicrystal's instability to a uniform phason strain. 
To improve the calculations, a stable quasicrystal should be
modeled and the fluctuations in small samples (due to variations
in tiling configurations and in decoration rules)
must be better understood. 
\rem{The stable quasicrystals probably have a better match between the
positions of the atoms and the potential well minima. 
The ``frustration'' model suggests this would
reduce $\Kpp$ even further, but we aren't sure since (as
noted in the text) our calculation in fact went the other way.}

\stars
We thank M.  Kraj\v{c}i for the use of his relaxation code,
J. Zou, A. Carlsson and R. Phillips for the AlMn potentials, and
M. Windisch and J. Hafner for the (Al,Cu)Li potentials. 
We are grateful, for discussion and comments, to
M. de Boisssieu, V. Elser, M. Widom, 
and especially to M.~Mihalkovi\v{c} who
originally developed our decoration/relaxation methods. 
This work was supported by U.S. Dept. of Energy grant
DE-FG02-89ER-45405.

\end{document}

\begin{table}
\caption{Caption of table. Example of table.}
\[
\begin{tabular}{ccccccc}
\hline
\multicolumn{1}{c}{1.2.95} & \multicolumn{1}{c}{1.7.95} & 
\multicolumn{1}{c}{1.1.96} & \multicolumn{1}{c}{1.7.96} & 
\multicolumn{1}{c}{1.1.97} & \multicolumn{1}{c}{1.7.97} & 31.12.97 \\ 
\hline
\multicolumn{1}{c}{$-10.000.000$} & \multicolumn{1}{c}{} &
\multicolumn{1}{c}{$65.400$} & 
\multicolumn{1}{c}{$65.400$} & \multicolumn{1}{c}{$65.400$} & 
\multicolumn{1}{c}{$65.400$} & $
\begin{array}{r}
65.400 \\ 
+1.111.800
\end{array}
$ \\ \hline
\end{tabular}
\]
\label{Tab1}
\end{table}

bbbbbbbbbb

\section{References}
Literature citations of periodicals \cite{HK}, \cite{S}, books \cite{LS},
\cite{LL}, conference proceedings \cite{GK} and preprints \cite{P}.
If you use EuroPhys.sty, you can use useful macros in the
bibliography environment; their use is described in this file.


\end{document}

%% file: euromacr.tex
\def\And{{\rm and\ }}

\def\stars{\bigskip\centerline{***}\medskip}

\newif\ifboo \boofalse

\def\Review#1{\boofalse{\it #1},}
\def\Name#1{{\sc #1},}
\def\Vol#1{\ifboo Vol. {\bf #1}\else{\bf #1}\fi}
\def\Year#1{\ifboo #1\else(#1)\fi}
\def\Book#1{\bootrue{\it #1},}
\def\Page#1{\ifboo {\rm p. #1}\else{\rm #1}\fi}